\documentclass[twocolumn,showpacs,prb,superscriptaddress]{revtex4}

\usepackage{graphicx}

\begin{document}

\title{Probing the Almeida-Thouless line away from the mean-field model}

\author{Helmut G.~Katzgraber}
\affiliation
{Theoretische Physik, ETH H\"onggerberg,
CH-8093 Z\"urich, Switzerland}

\author{A.~P.~Young}
\affiliation{Department of Physics,
University of California,
Santa Cruz, California 95064, USA}

\date{\today}

\begin{abstract}
Results of Monte Carlo simulations of the one-dimensional long-range
Ising spin glass with
power-law interactions in the presence of a (random) field are
presented. By tuning the exponent of the power-law interactions, we are able 
to scan the full range of possible behaviors from the infinite-range
(Sherrington-Kirkpatrick) model to the short-range model.  A finite-size
scaling analysis of the correlation length indicates that the Almeida-Thouless
line does not occur in the region with non-mean-field critical behavior in
zero field. However, there is evidence that an Almeida-Thouless line does
occur in the mean-field region.
\end{abstract}

\maketitle

\section{Introduction}
\label{sec:introduction}

The behavior of spin glasses in a magnetic field is still controversial.
While the infinite-range (mean-field)
Sherrington-Kirkpatrick (SK)
model\cite{sherrington:75} 
has a line of transitions at finite field known
as the Almeida-Thouless (AT) line,\cite{almeida:78} it has not been definitely
established 
whether an AT line occurs in more realistic models with
short range interactions. 
Previous numerical
studies\cite{bhatt:85,ciria:93b,kawashima:96,billoire:03b,marinari:98d,houdayer:99,krzakala:01,takayama:04,young:04}
have yielded conflicting results: some data
support the existence of an AT line in short-range spin glasses while others
claim its absence. Recently,\cite{young:04} a new approach using the
correlation length\cite{cooper:82,palassini:99b,ballesteros:00,martin:02} 
at {\em finite
fields} has been applied to the three-dimensional Edwards-Anderson Ising 
spin-glass model. The data of Ref.~\onlinecite{young:04}
indicate that, even for small fields,
there is no AT line in
three-dimensional spin glasses.

Here we use the techniques of Ref.~\onlinecite{young:04} to
study the crossover between mean-field models and
short-range spin glasses \textit{continuously}
by using a one-dimensional Ising chain with random power-law
interactions. The model's advantages are twofold: first, a large range of
system sizes can be studied, and second, by tuning the power-law exponent of
the interactions, the universality class of the model can be changed
continuously from the mean-field universality class to the short-range
universality class. We find that
there appears to be \textit{no} AT line
for the range of the power-law exponent corresponding to a non-mean-field
transition in zero field.
By analogy, this suggests
that there is also no AT line for short-range spin glasses, at least below the
upper critical dimension.

The paper is organized as follows. In Sec.~\ref{sec:details}, we introduce in
detail the model, observables, and numerical method used. In
Sec.~\ref{sec:results}, we present our results, and in
Sec.~\ref{sec:conclusions}, we summarize our findings.

\section{Model, Observables, and Numerical Details}
\label{sec:details}

The Hamiltonian of the one-dimensional long-range Ising spin glass
with random power-law interactions\cite{bray:86b,fisher:88} is given by
\begin{equation}
{\cal H} = -\sum_{\langle i,j \rangle} J_{ij} S_i S_j - \sum_i h_i S_i\; ,
\label{eq:hamiltonian}
\end{equation}
where $S_i = \pm 1$ represents Ising spins evenly
distributed on a ring of length $L$ in order to ensure periodic boundary
conditions. The sum is over all spins on the chain and the couplings $J_{ij}$
are given by\cite{katzgraber:03}
\begin{equation}
J_{ij} = c(\sigma)\frac{\epsilon_{ij}}{r_{ij}^\sigma}\; ,
\label{eq:bonds}
\end{equation}
where the $\epsilon_{ij}$ are chosen according to a Gaussian distribution
with zero mean and standard deviation unity,
and $r_{ij} = (L/\pi)\sin[(\pi |i - j|)/L]$ represents the
{\em geometric} distance between the spins on
the ring.\cite{distances}
The power-law exponent $\sigma$ characterizes the
interactions and, hence, determines the universality class of the model.
The constant $c(\sigma)$ in Eq.~(\ref{eq:bonds}) is chosen to give a
mean-field
transition temperature $T_{\rm c}^{\rm MF} = 1$, where
\begin{equation}
\left(T_{\rm c}^{\rm MF}\right)^2 = \sum_{j\ne i} [ J_{ij}^2]_{\rm av} =
c(\sigma)^2 \sum_{j\ne i} \frac{1}{r_{ij}^{2\sigma}} \; .
\label{eq:tcmf}
\end{equation}
Here $[\cdots]_{\rm av}$ denotes an average over disorder.

\begin{table}[!tb]
\caption{
A summary of the behavior for different ranges of $\sigma$ in one space
dimension and at zero field. 
IR means infinite
range, i.e., $\sum_{j\ne i} J_{ij}^2$ diverges unless the bonds $J_{ij}$ 
are scaled by an inverse power of the system size. LR means that the behavior 
is dominated by the long-range tail of the interactions, and SR means that 
the behavior is that of a short-range system.
\label{tab:ranges}
}
\begin{tabular*}{\columnwidth}{@{\extracolsep{\fill}} l l }
\hline
\hline
$\sigma$ & behavior  \\
\hline
$\sigma = 0$ & SK model \\
$0 < \sigma \le 1/2$ & IR  \\
$1/2 < \sigma < 2/3$ & LR (mean field with $T_{\rm c} > 0$) \\
$2/3 < \sigma \le 1$ & LR (non-mean field with $T_{\rm c} > 0$) \\
$1 < \sigma \le 2$ & LR ($T_{\rm c} = 0$) \\
$\sigma \ge 2$ & SR ($T_{\rm c} = 0$) \\
\hline
\hline
\end{tabular*}
\end{table}

In Eq.~(\ref{eq:hamiltonian}), the spins couple to site-dependent random 
fields $h_i$ chosen from a Gaussian
distribution with zero mean $[h_i]_{\rm av} = 0$ and standard deviation 
$[h_i^2]_{\rm av}^{1/2} = H_{\rm R}$. For a symmetric distribution of bonds, 
the sign of $h_i$ can be ``gauged away'' so a uniform field is completely 
equivalent to a bimodal distribution of fields with $h_i = \pm H_{\rm R}$. 
While the AT line is usually studied for the case of a uniform field,
the SK model with Gaussian 
random fields (as considered here)
also shows an AT line. For short-range three-dimensional spin glasses it has
been shown in Ref.~\onlinecite{young:04} that results for 
Gaussian-distributed random fields agree within error bars with results for a 
uniform field. The use of Gaussian-distributed random fields has the advantage 
over a uniform external field in that we can apply a useful equilibration
test,\cite{katzgraber:01,katzgraber:03,young:04} see Eq.~(\ref{eq:U}) below.

From Eq.~(\ref{eq:tcmf}), we see that for $\sigma \le 1/2,\ 
c(\sigma)$ varies with a power of the system, $c(\sigma) \sim
L^{-(1-2\sigma)/2}$, for large $L$. We shall denote systems in this region as
``infinite range'' (IR). The extreme limit of $\sigma = 0$ gives the SK model,
whose solution is the mean-field (MF) theory for spin glasses.  For
$\sigma > 1/2,\ c(\sigma)$ tends to a constant as $L \to\infty$. As discussed
in an earlier work (see Ref.~\onlinecite{katzgraber:03} and references 
therein),
for $1/2 < \sigma \le 1$, the system has a finite-temperature transition into a
spin-glass phase in a long-range (LR) universality class at zero field.
For $1 < \sigma \le
2$, the system has $T_{\rm c} = 0$ 
and the critical behavior is also determined by
the LR universality class. For $\sigma > 2$, we have a short-range (SR)
universality class with $T_{\rm c} = 0$. 
Finally, we note\cite{kotliar:83} that for
$1/2 < \sigma < 2/3$ the critical behavior is mean-field-like, while for
$2/3 < \sigma \le 1$ it is non-mean field like.
This behavior is summarized in Table \ref{tab:ranges}.
Critical exponents depend continuously on
$\sigma$ in the LR regime,
but are independent of $\sigma$ in the SR region.  Here
we focus on the regime $1/2 < \sigma \le 1$ because there
the system exhibits a finite-temperature transition that can be
tuned continuously away from the
mean-field universality limit by changing the exponent $\sigma$.

To determine the existence of an AT line, we compute the two-point correlation
length.\cite{palassini:99b,ballesteros:00,young:04} 
We calculate the wave-vector-dependent
spin-glass susceptibility which is defined by
\begin{equation}
\chi_{\rm SG}(\mathbf{k}) = \frac{1}{N} \sum_{i, j} \left[\Big(
\langle S_i S_j\rangle_T - \langle S_i \rangle_T \langle S_j\rangle_T
\Big)^2 \right]_{\rm av}\!\!\!\!\! e^{i\mathbf{k}\cdot(\mathbf{R}_i -
\mathbf{R}_j)} ,
\label{eq:chisg}
\end{equation}
where $\langle \cdots \rangle_T$ denotes a thermal average. 
Note that at zero
field $\langle S_i \rangle_T$ can be set to zero.
The correlation length of the finite system is then given by
\begin{equation}
\xi_L = \frac{1}{2 \sin (k_\mathrm{min}/2)}
\left[\frac{\chi_{\rm SG}(0)}{\chi_{\rm SG}({\bf k}_\mathrm{min})} -
1\right]^{1/(2\sigma-1)},
\label{xiL}
\end{equation}
where ${\bf k}_\mathrm{min} = (2\pi/L, 0, 0)$ is the smallest nonzero
wave vector.
The reason for the power $1/(2\sigma-1)$ is that at long wavelengths,
we expect a \textit{modified} Ornstein-Zernicke form\cite{cooper:82,martin:02}
\begin{equation}
\chi_{\rm SG}({\bf k}) \propto \left(v + k^{2\sigma - 1}\right)^{-1}
\end{equation}
for the long-range case, where $v$ is a measure of the deviation from 
criticality. It follows that the bulk correlation length $\xi$ diverges 
for $v \to 0$ like $v^{-1/(2\sigma-1)}$.

\begin{figure}
\includegraphics[width=\columnwidth]{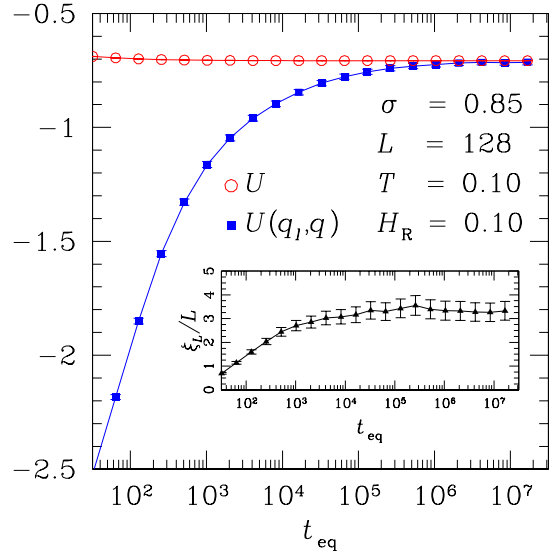}
\vspace*{-1.0cm}
\caption{(Color online)
Sample equilibration plot for $\sigma = 0.85$, $L = 128$, $H_{\rm R} = 0.1$,
and $T = 0.10$ (the lowest temperature simulated at finite fields).
Data for the average energy $U$, and
$U(q_l, q)$ defined in Eq.~(\ref{eq:U}),
as a function of equilibration time
$t_{\rm eq}$. They approach their common value
from opposite directions and, once they agree, do not change on further
increasing $t_{\rm eq}$.
The inset shows data for the
correlation length divided by system size as a function of equilibration time.
The data are independent of $t_{\rm eq}$ once $U$ and
$U(q_l, q)$ agree.
}
\label{fig:equil}
\end{figure}
                                                                                
\begin{table}[!tb]
\caption{
Parameters of the simulations for $H_{\rm R} = 0.0$. $N_{\rm sa}$ 
is the number of samples, $N_{\rm sw}$ is the total number of Monte Carlo 
sweeps for each of the $2 N_T$ replicas for a single sample, 
$T_{\rm min}$ is the lowest temperature simulated, and $N_T$ is the number 
of temperatures used in the parallel tempering method for each system size 
$L$ and power-law exponent $\sigma$.
\label{tab:simparams0}
}
\begin{tabular*}{\columnwidth}{@{\extracolsep{\fill}} c r r r r r }
\hline
\hline
$\sigma$ & $L$  &  $N_{\rm sa}$  & $N_{\rm sw}$ & $T_{\rm min}$ & $N_{T}$  \\
\hline
$0.55$ &  $32$ &  $5000$ &  $10240$ & $0.405$ & $15$ \\
$0.55$ &  $64$ &  $5000$ &  $10240$ & $0.405$ & $15$ \\
$0.55$ & $128$ &  $5000$ &  $20480$ & $0.405$ & $15$ \\
$0.55$ & $256$ &  $5000$ & $102400$ & $0.405$ & $15$ \\
$0.55$ & $512$ &  $5000$ &  $32768$ & $0.630$ & $11$ \\[2mm] 

$0.65$ &  $32$ &  $5000$ &  $10240$ & $0.405$ & $15$ \\
$0.65$ &  $64$ &  $5000$ &  $10240$ & $0.405$ & $15$ \\
$0.65$ & $128$ &  $5000$ &  $20480$ & $0.405$ & $15$ \\
$0.65$ & $256$ &  $5000$ & $102400$ & $0.405$ & $15$ \\
$0.65$ & $512$ &  $5000$ & $524288$ & $0.405$ & $15$ \\[2mm]

$0.75$ &  $32$ &  $5000$ &  $10240$ & $0.405$ & $15$ \\
$0.75$ &  $64$ &  $5000$ &  $10240$ & $0.405$ & $15$ \\
$0.75$ & $128$ &  $5000$ &  $20480$ & $0.405$ & $15$ \\
$0.75$ & $256$ &  $5000$ & $102400$ & $0.405$ & $15$ \\
$0.75$ & $512$ &  $2500$ & $524288$ & $0.405$ & $15$ \\[2mm]

$0.85$ &  $32$ &  $5000$ &  $10240$ & $0.405$ & $15$ \\
$0.85$ &  $64$ &  $5000$ &  $20480$ & $0.405$ & $15$ \\
$0.85$ & $128$ &  $5000$ & $102400$ & $0.405$ & $15$ \\
$0.85$ & $256$ &  $5000$ & $204800$ & $0.405$ & $15$ \\
$0.85$ & $512$ &  $2500$ & $204800$ & $0.405$ & $15$ \\[2mm]
\hline
\hline
\end{tabular*}
\end{table}
\begin{table}[!tb]
\caption{
Parameters of the simulations for $H_{\rm R} = 0.1$. $N_{\rm sa}$
is the number of samples, $N_{\rm sw}$ is the total number of Monte Carlo
sweeps for each of the $4 N_T$ replicas for a single sample, $T_{\rm min}$ is
the lowest temperature simulated, and $N_T$ is the number of temperatures used
in the parallel tempering method for each system size $L$ and power-law
exponent $\sigma$. \label{tab:simparams}
}
\begin{tabular*}{\columnwidth}{@{\extracolsep{\fill}} c r r r r r }
\hline
\hline
$\sigma$ & $L$  &  $N_{\rm sa}$  & $N_{\rm sw}$ & $T_{\rm min}$ & $N_{T}$
\\
\hline
$0.55$ &  $32$ &  $5000$ &    $81920$ & $0.100$ & $26$ \\
$0.55$ &  $64$ &  $5000$ &   $327680$ & $0.100$ & $26$ \\
$0.55$ & $128$ &  $5000$ &  $1310720$ & $0.100$ & $26$ \\
$0.55$ & $256$ &  $2000$ &  $1048576$ & $0.405$ & $15$ \\
$0.55$ & $512$ &  $2000$ &    $65536$ & $0.760$ &  $9$ \\[2mm]

$0.65$ &  $32$ &  $5000$ &    $81920$ & $0.100$ & $26$ \\
$0.65$ &  $64$ &  $5000$ &   $327680$ & $0.100$ & $26$ \\
$0.65$ & $128$ &  $5000$ &  $1310720$ & $0.100$ & $26$ \\
$0.65$ & $256$ &  $2000$ &  $1048576$ & $0.195$ & $20$ \\
$0.65$ & $512$ &  $2000$ &   $524288$ & $0.500$ & $13$ \\[2mm]

$0.75$ &  $32$ &  $5000$ &    $81920$ & $0.100$ & $26$ \\
$0.75$ &  $64$ &  $5000$ &   $327680$ & $0.100$ & $26$ \\
$0.75$ & $128$ &  $5000$ &  $1310720$ & $0.100$ & $26$ \\
$0.75$ & $256$ &  $2000$ &  $8388608$ & $0.100$ & $26$ \\[2mm]

$0.85$ &  $32$ &  $5000$ &    $81920$ & $0.100$ & $26$ \\
$0.85$ &  $64$ &  $5000$ &   $327680$ & $0.100$ & $26$ \\
$0.85$ & $128$ &  $2000$ & $16777216$ & $0.100$ & $26$ \\[2mm]
\hline
\hline
\end{tabular*}
\end{table}

The correlation length divided by the system size $\xi_L/L$ has
the following scaling property:
\begin{equation}
\frac{\xi_L}{L} = \widetilde{X}\left(L^{1/\nu}[T - T_{\rm c}(H_{\rm
R})]\right) \; ,
\label{fss}
\end{equation}
where $\nu$ is the correlation length exponent and $T_{\rm c}(H_{\rm R})$ is
the transition temperature for a field of strength $H_{\rm R}$. This
behavior is similar
to that of the
Binder ratio,\cite{binder:81} 
but it shows a clearer signature of the transition as the data are not 
restricted to a finite interval. 

In order to test equilibration of the Monte Carlo method, we also compute the
link overlap\cite{katzgraber:03} $q_{l}$ given by
\begin{equation}
q_l = \frac{2}{N}\sum_{\langle i,j\rangle}
\frac{[J_{ij}^2]_{\rm av}}{(T_{\rm c}^{MF})^2}
[ 
\langle S_i^\alpha S_j^\alpha S_i^\beta  S_j^\beta  \rangle_T
]_{\rm av} \; ,
\label{eq:ql}
\end{equation}
where $T_{\rm c}^{\rm MF}$ is given by Eq.~(\ref{eq:tcmf}) and $\alpha$ and
$\beta$ refer to two replicas of the system with the same disorder.
In addition, we compute the spin overlap $q$ given by
\begin{equation}
q = \frac{1}{N}\sum_{i = 1}^{N} 
[ \langle S_i^\alpha S_i^\beta \rangle_T ]_{\rm av} .
\label{eq:q}
\end{equation}

Because both the fields and interactions have a Gaussian distribution,
integrating by parts the expression for the average energy per spin $U$
gives\cite{young:04,katzgraber:03,katzgraber:01}
\begin{equation}
U \equiv U(q_l, q) =  -{(T_{\rm c}^{\rm MF})^2 \over 2 T} ( 1 - q_l) 
- {H_{\rm R}^2 \over T} (1 - q) \, .
\label{eq:U}
\end{equation}

\begin{figure*}[!tb]
\includegraphics[width=\columnwidth]{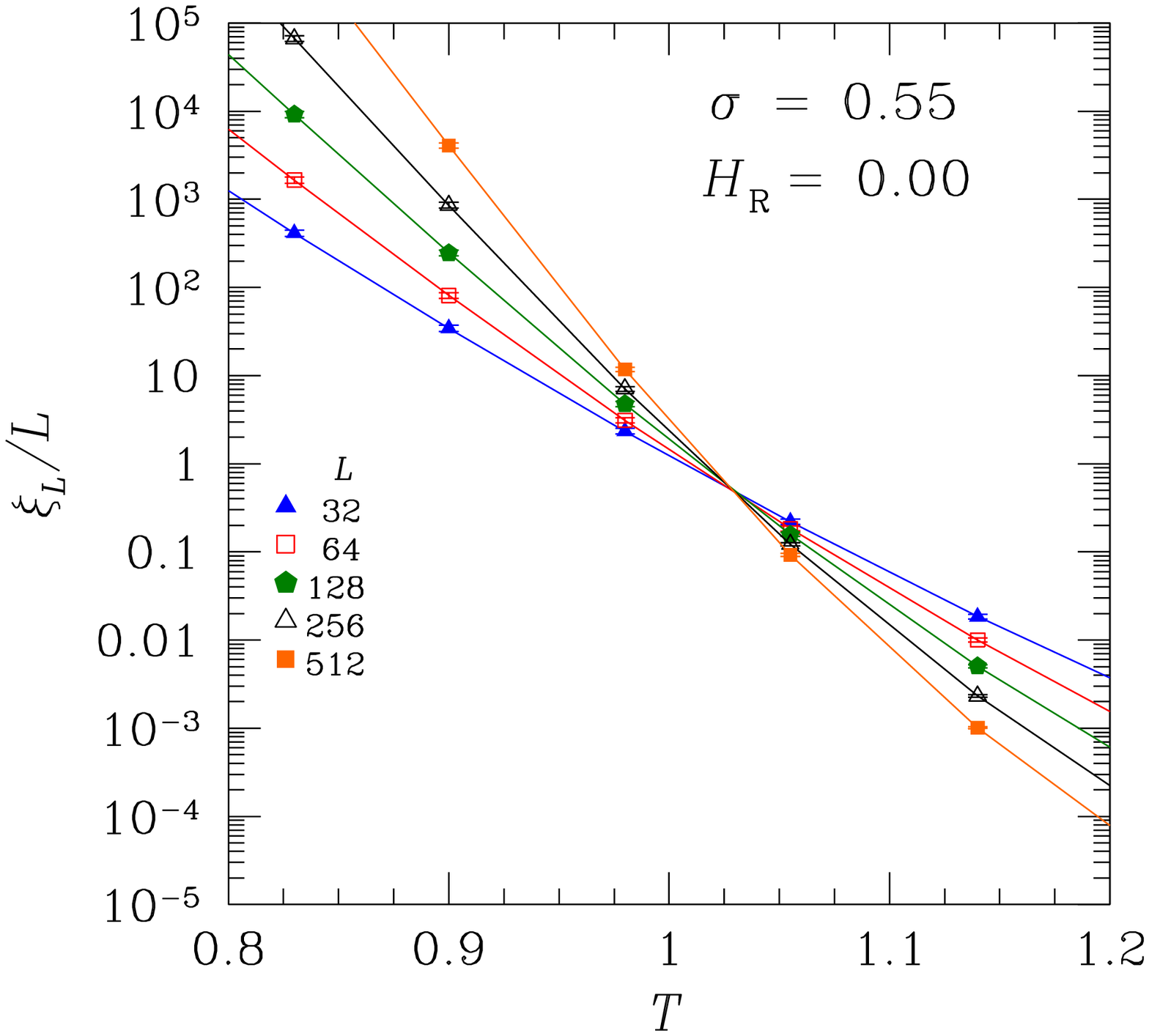}
\includegraphics[width=\columnwidth]{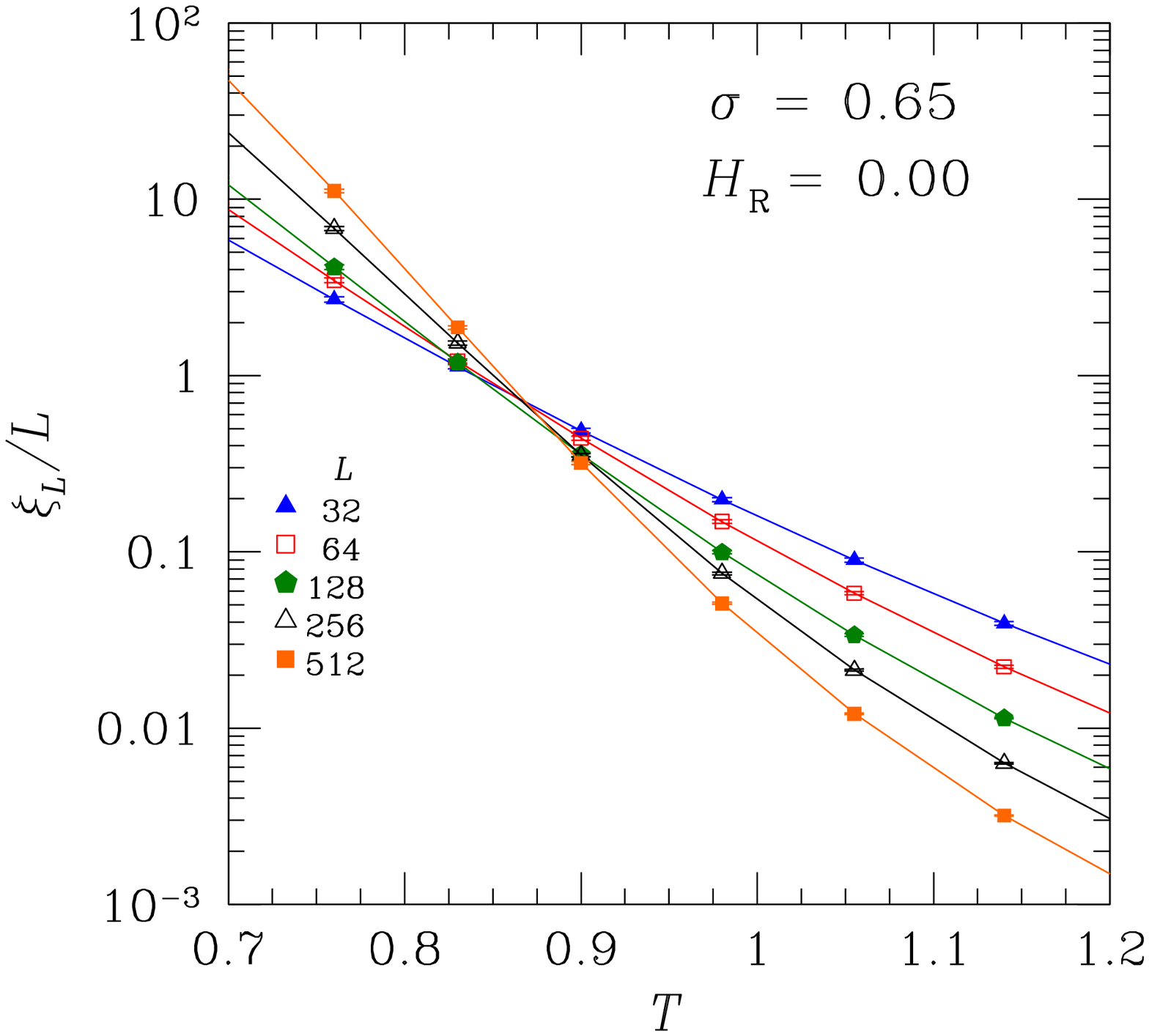}
\vspace*{-1.2cm}

\includegraphics[width=\columnwidth]{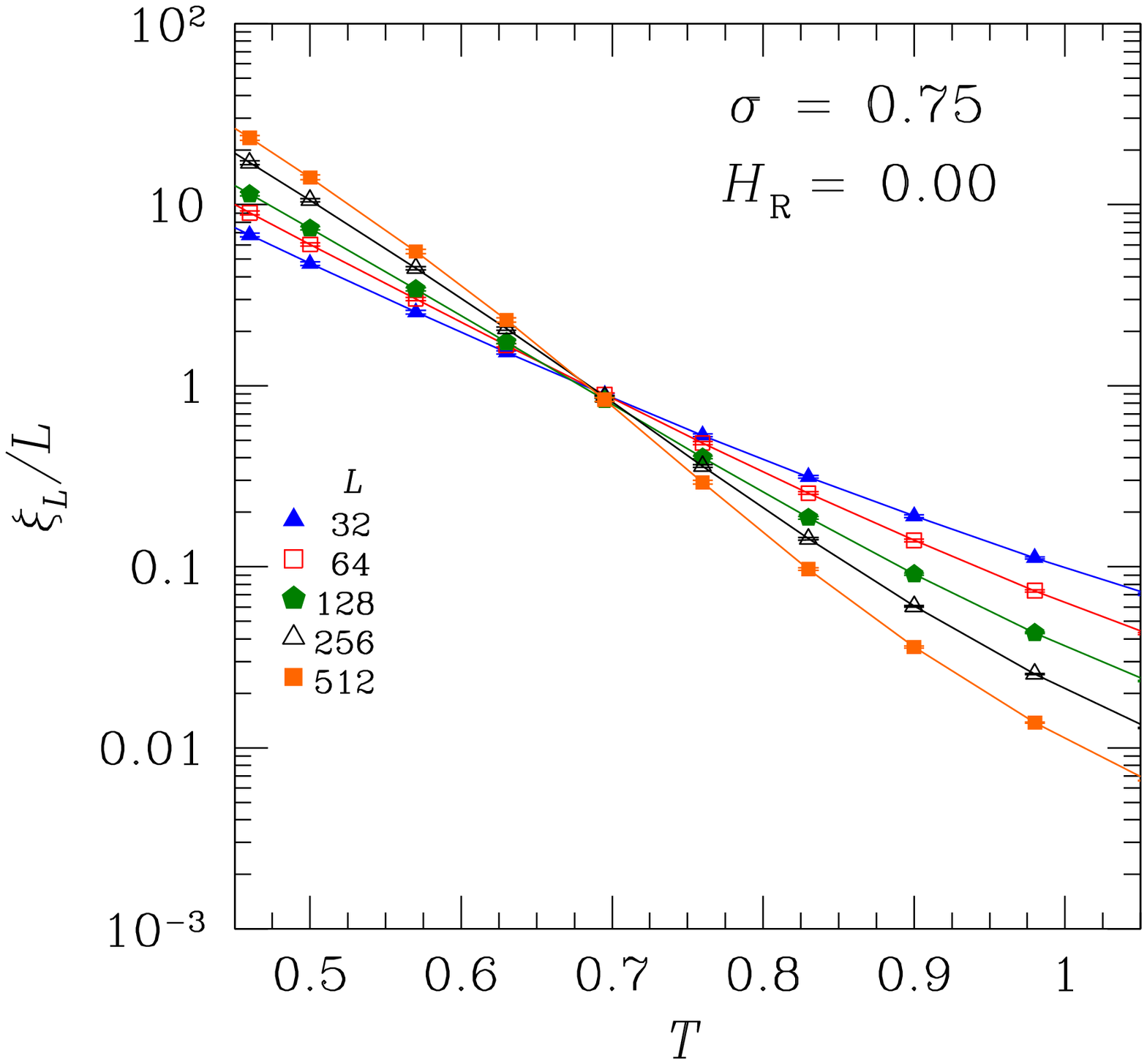}
\includegraphics[width=\columnwidth]{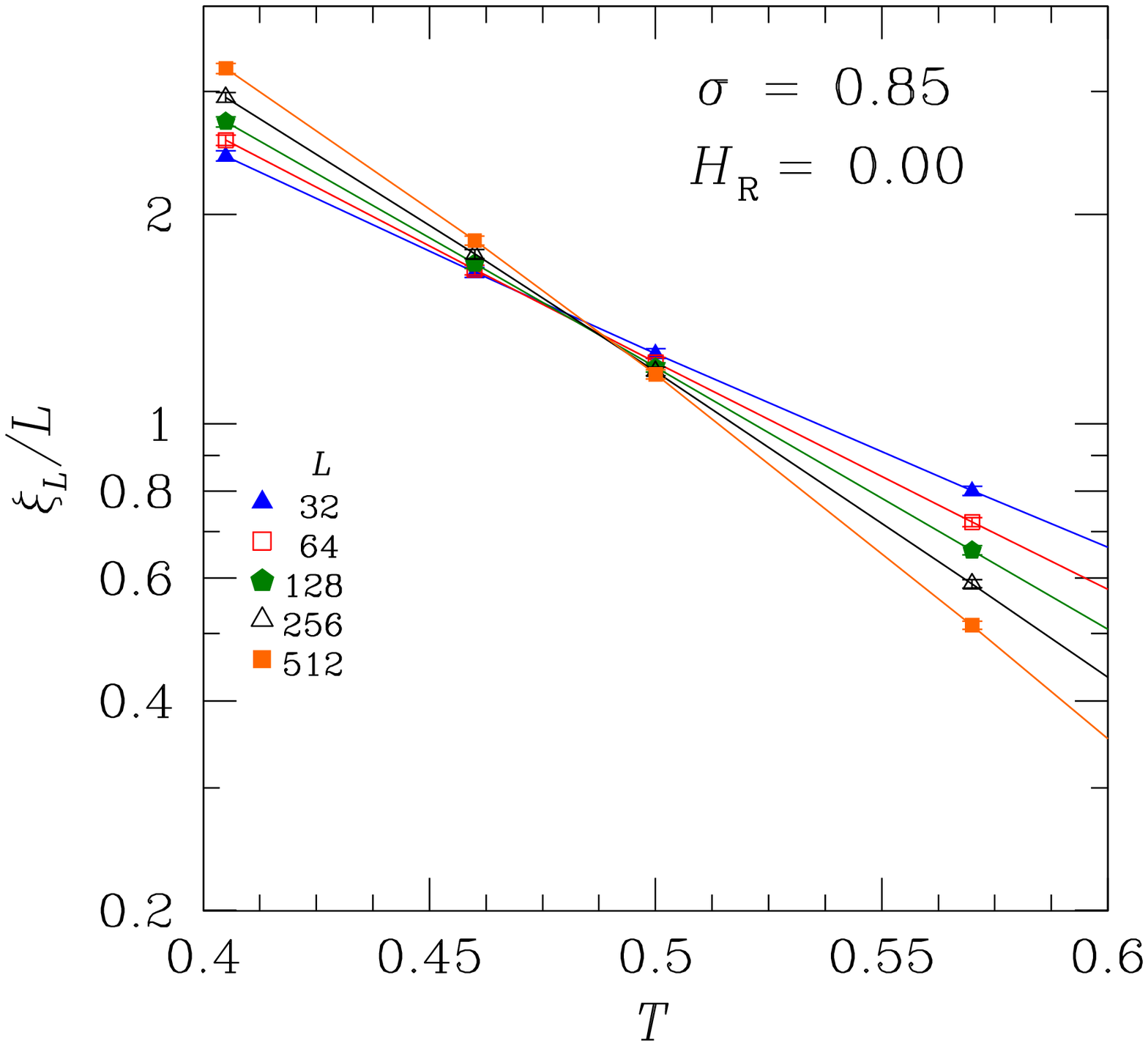}
\vspace*{-1.0cm}

\caption{(Color online)
Each figure shows data for $\xi_L/L$ vs $T$ for $H_{\rm R}=0$ for different
system sizes, for a particular value of $\sigma$.
For all values of $\sigma$, the data cross indicating that there is a spin-glass
transition at finite temperature.
}
\label{fig:heq0}
\end{figure*}

As shown in Fig.~\ref{fig:equil},
when starting from a random spin configuration,
$U$ approaches its equilibrium value
from above while $U(q_l, q)$ approaches its equilibrium value from below.
Once $U = U(q_l, q)$, the data do not change by further
increasing the number of
Monte Carlo steps, which shows that the system is
in equilibrium.
It is
also important to ensure that other observables are also
in equilibrium once $U = U(q_l, q)$, and this is
shown in the inset to Fig.~\ref{fig:equil} for the case of the correlation
length.

The simulations are done using the parallel tempering Monte Carlo
method.\cite{hukushima:96,marinari:96} The method is not as efficient in a
field,\cite{moreno:03,billoire:03} but nevertheless it performs considerably
better than simple Monte Carlo.
In order to compute the products of up to four thermal averages in 
Eq.~(\ref{eq:chisg}) without bias, we simulate four
copies (replicas) of the system with the same bonds and fields at each
temperature. Simulations are performed at zero field, as well as at 
$H_{\rm R} = 0.1$, a field that is considerably smaller than the
$\sigma$-dependent transition temperature $T_{\rm c}$.
Parameters of the
simulations at zero and finite fields are presented in Tables 
\ref{tab:simparams0} and \ref{tab:simparams}, respectively.

\section{Results}
\label{sec:results}

We first consider the case of zero field and take
$\sigma = 0.55$, $0.65$, $0.75$, and $0.85$. The values $\sigma = 0.75$ and
$0.85$ are in the non-MF region (see Table \ref{tab:ranges}) while
$\sigma = 0.55$ is in the MF region and, furthermore, is close to the
value ($\sigma = 1/2$) where the system becomes infinite range. The value
$\sigma = 0.65$ is close to the point $\sigma = 2/3$ where the critical
behavior changes from MF to non-MF.
The data are shown in
Fig.~\ref{fig:heq0}.
In all cases, the data cross 
at a transition temperature which we determine as $T_{\rm c} = 1.03(3)$ 
for $\sigma = 0.55$, $0.86(2)$ for $\sigma=0.65$,
$0.69(1)$ for $\sigma = 0.75$, and $0.49(1)$ for $\sigma = 0.85$.
Note that $T_{\rm c}$ decreases continuously with increasing $\sigma$ and is
expected to drop 
to zero at $\sigma = 1$.\cite{katzgraber:03} For the SK model ($\sigma = 0$),
one has $T_{\rm c} = 1$, essentially the result we find for $\sigma = 0.55$,
so it is possible that $T_{\rm c}$ has little variation with $\sigma$
for $\sigma \le 0.55$. 

\begin{figure*}[!tb]
\includegraphics[width=\columnwidth]{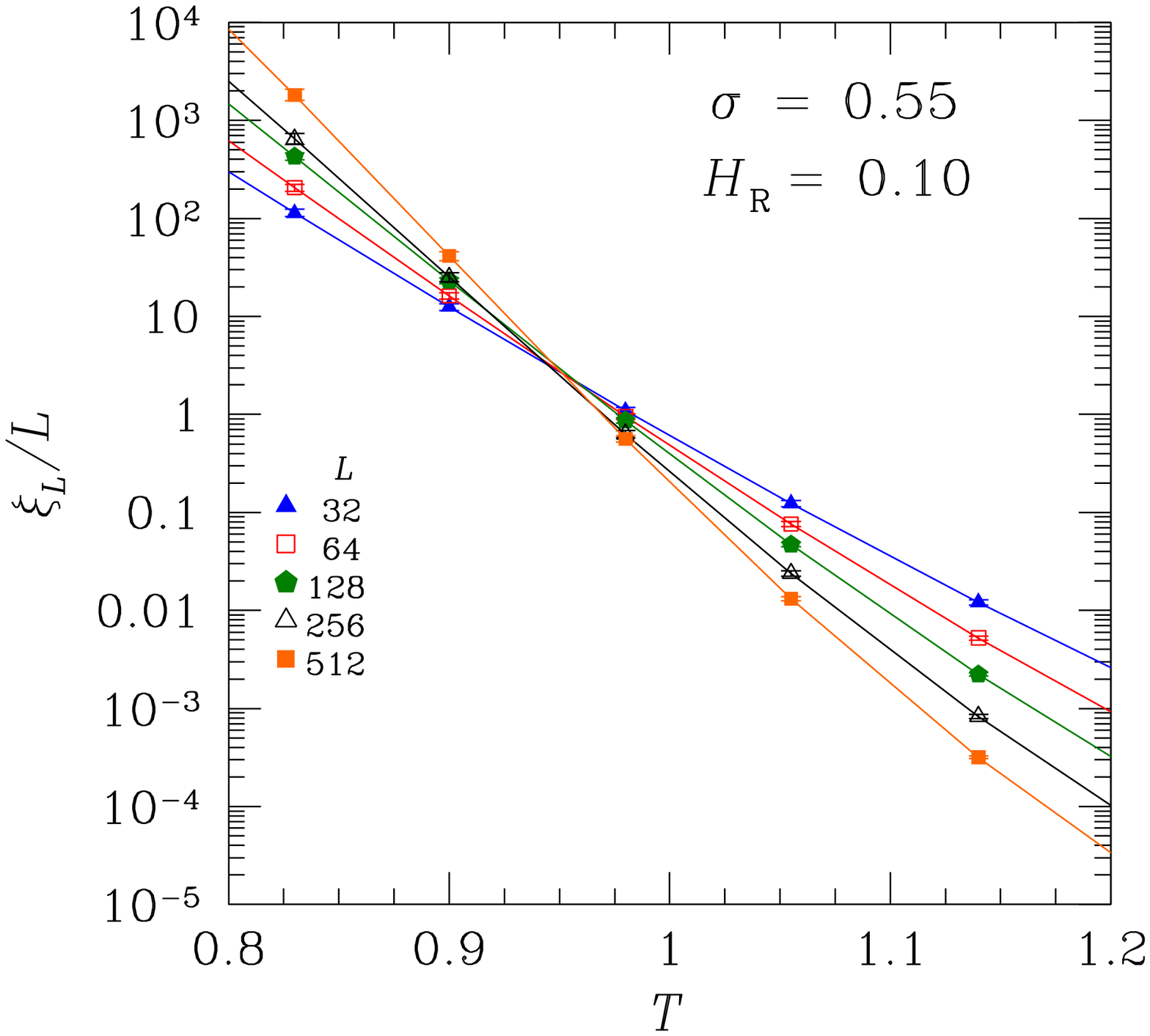}
\includegraphics[width=\columnwidth]{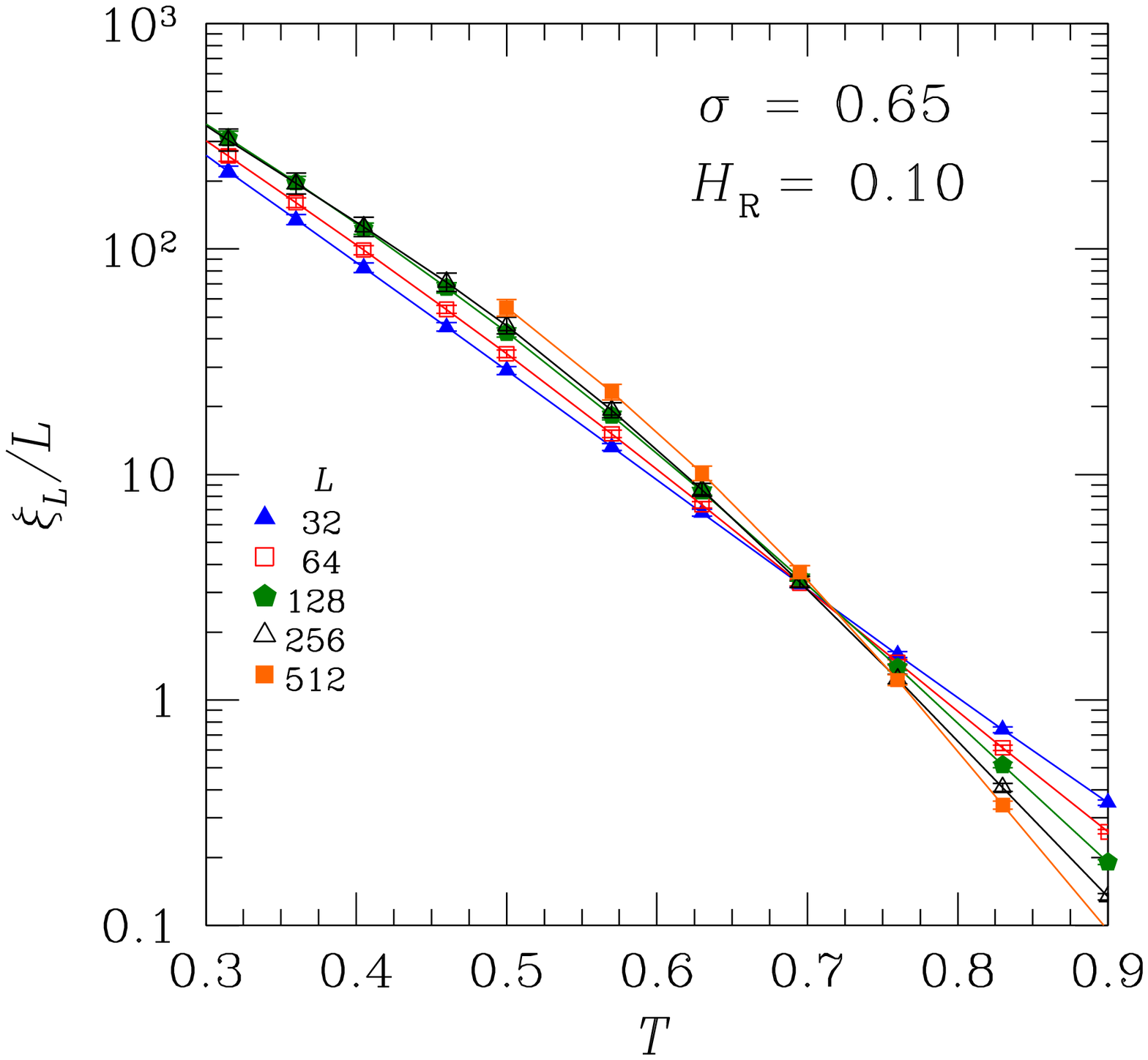}
\vspace*{-1.2cm}

\includegraphics[width=\columnwidth]{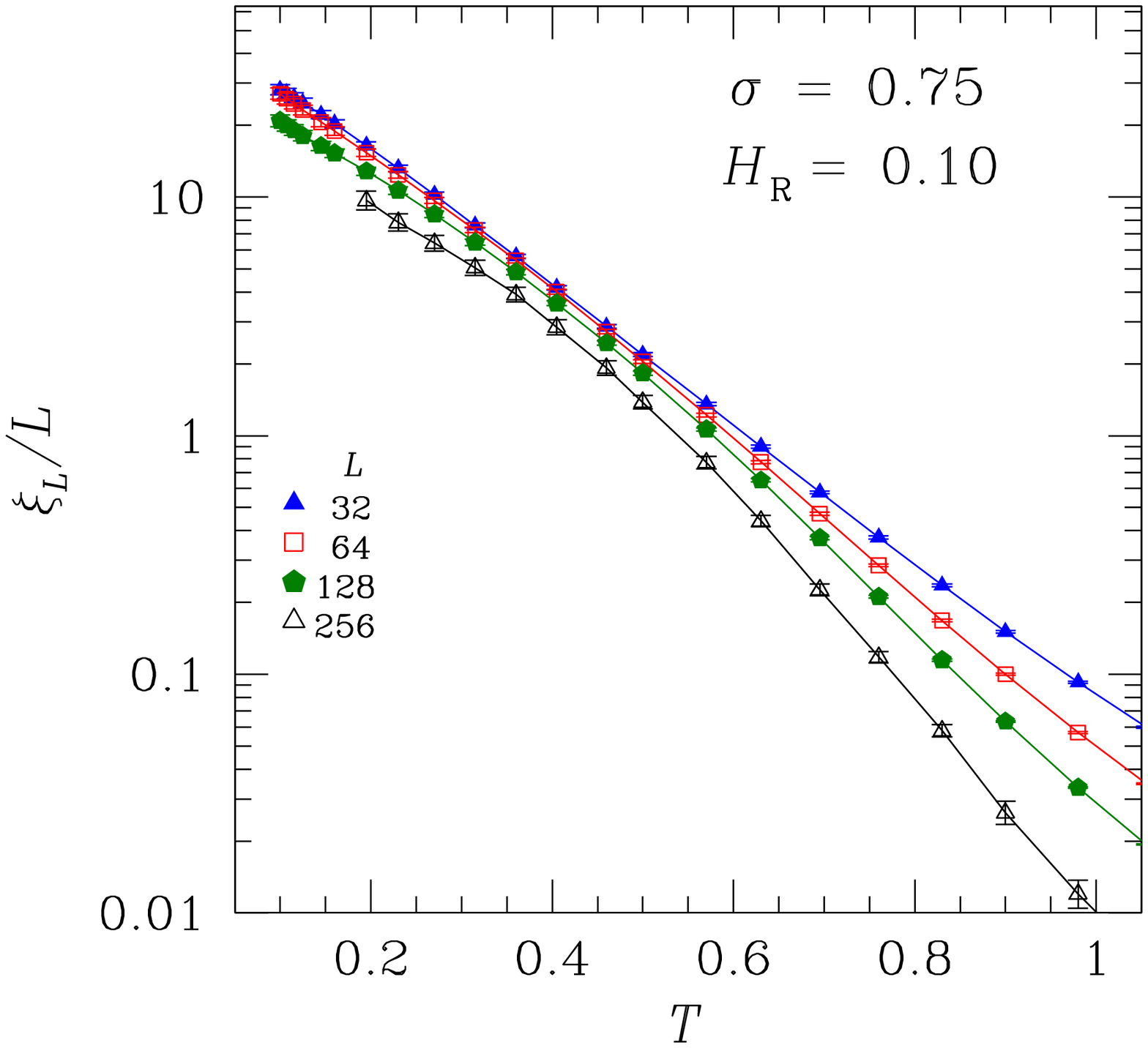}
\includegraphics[width=\columnwidth]{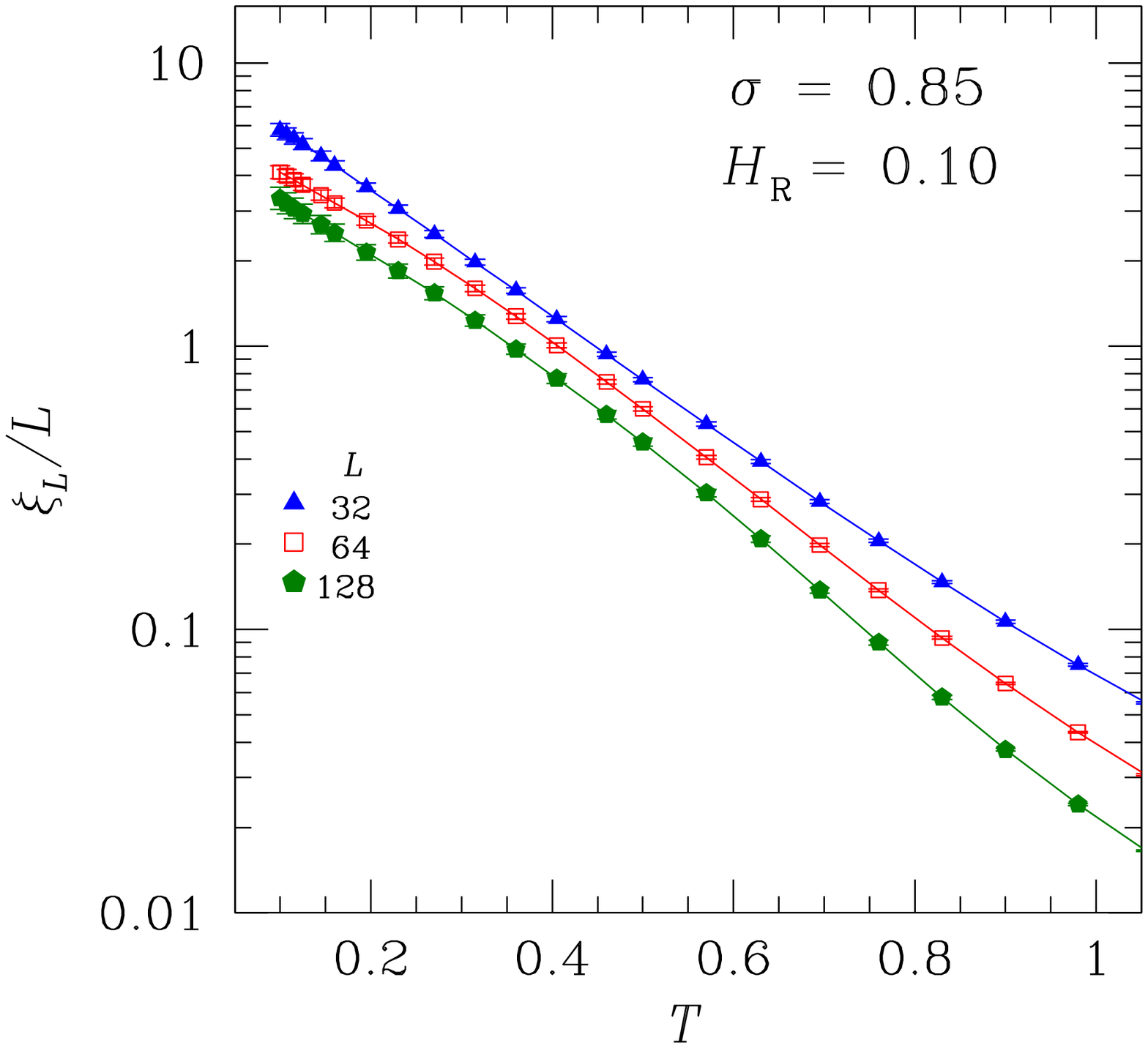}
\vspace*{-1.0cm}

\caption{(Color online)
Each figure shows results for $\xi_L/L$ vs $T$ for $H_{\rm R}=0.10$ for 
different system sizes, for a particular 
value of $\sigma$.
For $\sigma = 0.55$, the data cross at $T_{\rm c} = 0.96(2)$
showing that an AT line seems to be present. 
For $\sigma = 0.65$, the data show close to marginal behavior. 
This may indicate that
$\sigma = 0.65$ is close to the borderline value for having an AT line.
For $\sigma = 0.75$ and $\sigma = 0.85$, the data do not cross for any 
temperature down to $T = 0.10$, which is considerably smaller than the zero 
field transition temperatures. This indicates that there is no AT line.
Overall the results indicate that, on increasing $\sigma$, the AT line 
disappears.
}
\label{fig:hgt0}
\end{figure*}

Next we consider $H_{\rm R} = 0.10$ and show the data in
Fig.~\ref{fig:hgt0}. The results for $\sigma = 0.75$ and $0.85$, which are
in the non-mean-field regime, show no sign of a transition. However,
the data for $\sigma = 0.55$ do show a signature of a transition at $T_{\rm
c} = 0.96(2) < T_{\rm c}(H_{\rm R} = 0)$. Whether this would persist up to
infinite system sizes is not clear, but it certainly cannot be ruled out.
The results for $\sigma = 0.55$ show that the method used here \textit{is} 
capable of detecting an AT line in the presence of a field.
For $\sigma = 0.65$, the data shows a marginal behavior.
Since $\sigma = 0.65$ is close to the value of $2/3$ which
separates MF and non-MF behavior in zero field, this marginal behavior may 
indicate that $2/3$ is also the borderline value below which an AT line occurs.
An alternative possibility, which
we cannot rule out, is that an AT line only occurs in the infinite-range
region ($\sigma < 1/2$) but that as $\sigma$ is decreased toward $1/2$,
one needs to study larger system
sizes to see the absence of a transition.

\section{Conclusions}
\label{sec:conclusions}
We have considered
a one-dimensional spin-glass model with long-range interactions
that allows 
the universality class to be changed from
the infinite-range limit to the short-range case by tuning the power-law
exponent $\sigma$ of the interactions. We find that
there does not appear to be an AT line in a field for models with $\sigma$ in
the range where there is
non-mean-field critical behavior at zero field. However, in the
region of $\sigma$ that is not infinite-range but has mean-field critical
behavior
($1/2 < \sigma < 2/3$), there does appear to be an AT line. 

These conclusions rely on extrapolating from finite sizes to the thermodynamic
limit. It would be
particularly interesting to know if the conclusion that there is an AT line
for $\sigma = 0.55$ persists in the thermodynamic limit, or whether the AT
line really only occurs in the infinite-range case ($\sigma < 1/2$). It is
possible that as $\sigma$ is decreased, larger sizes or larger values of 
$H_{\rm R}$ are needed to probe the asymptotic behavior. 
Therefore, it would be desirable
to simulate a range of values of $H_{\rm R}$, especially for $\sigma = 0.55$.
We have some results for $H_{\rm R} = 0.2$ for a relatively small number of
samples and for sizes only up to $L=96$,
which indicate a crossing at a lower temperature than for $H_{\rm R} =
0.1$. However, we are unable to carry out a systematic study of the dependence
on $H_{\rm R}$ because
the results presented above already required considerable computer
time, and the parallel tempering algorithm becomes less efficient at larger
fields.

Making an analogy between the one-dimensional 
long-range model for different values of
$\sigma$ and short-range models for different values of space dimension $d$,
we infer that there is no AT line for short-range spin glasses in the
non-mean-field regime, i.e., below the upper critical dimension $d_{\rm u}=6$.
However, there
may be an AT line above the upper critical dimension. Speculations along these
lines have also been made very recently by Moore.\cite{moore:05}

\begin{acknowledgments}
A.~P.~Y.~acknowledges support from the National Science Foundation 
under NSF Grant No.~DMR 0337049.
The simulations were performed on the Hreidar and Gonzales clusters
at ETH Z\"urich. We thank M.~A.~Moore for helpful discussions and suggestions.
\end{acknowledgments}

\bibliography{comment,refs}

\begin{thebibliography}{27}
\bibitem[{dis()}]{distances}
\bibinfo{note}{Note that the distance $r_{ij}$ between two spins $i$ and $j$ on
  the chain is determined by $r_{ij} = 2R\sin(\alpha/2)$, where $R$ is the
  radius of the chain and $\alpha$ is the angle between the two sites on the
  circle. The previous expression can be rewritten in terms of the system size
  $L$ and the positions of the spins to obtain $r_{ij} = (L/\pi)\sin\left(\pi
  |i - j|/L\right)$.}
\end{thebibliography}

\end{document}